# Programmable In-Network Obfuscation of Traffic

Liang Wang, Hyojoon Kim, Prateek Mittal, Jennifer Rexford
Princeton University

## ABSTRACT

Recent advances in programmable switch hardware offer a fresh opportunity to protect user privacy. This paper presents PINOT, a lightweight in-network anonymity solution that runs at line rate within the memory and processing constraints of hardware switches. PINOT encrypts a client's IPv4 address with an efficient encryption scheme to hide the address from downstream ASes and the destination server. PINOT is readily deployable, requiring no end-user software or cooperation from networks other than the trusted network where it runs. We implement a PINOT prototype on the Barefoot Tofino switch, deploy PINOT in a campus network, and present results on protecting user identity against public DNS, NTP, and WireGuard VPN services.

## 1 INTRODUCTION

Network traffic contains privacy-sensitive information. While encryption protocols such as TLS provide confidentiality of data, they do not hide sensitive meta-data such as the identities of endpoints. Specifically, an Internet Protocol (IP) address can still be used to pinpoint and identify a user and device communicating on the Internet, putting privacy and security at risk [17–19, 22, 31, 43, 45]. However, existing approaches for anonymous communications either introduce high performance overhead (e.g., Tor) [12, 42] or face significant deployment challenges [4–6, 11, 24, 37].

Programmable switch hardware creates an opportunity to build a high-performance anonymity system by offloading privacy functionality to the network. Nevertheless, programmable switch hardware has limited memory and processing resources, posing challenges for implementing cryptographic algorithms that are commonly used in privacy applications. This begs the question: can we leverage programmable data planes to design a readily-deployable anonymity system that balances the privacy/performance trade-off?

In this paper, we push the boundaries of offloading privacy functionality to programmable data planes. We present PINOT (Programmable In-Network Obfuscation of Traffic), a lightweight anonymity system that runs in programmable switch hardware. In this preliminary work, we focus on protecting client IPv4 addresses against particular public services that provide DNS, NTP, and WireGuard VPN (a connectionless VPN) services, and will extend PINOT to support more services in future work.

PINOT runs at the border of a trusted network and encrypts the users' IPv4 addresses in packet headers before packets leave the network, as shown in Figure 1. We use a secure and efficient cipher built atop iterated Even-Mansour (EM) [2], which can perform encryption and decryption in a manner that is compatible with a single pass through the packet-processing pipeline of a hardware switch. Thus, PINOT can encrypt/decrypt IP addresses at hardware switch rates (e.g., up to 12.8 Tbps on a state-of-the-art Tofino switch). Leveraging the growing IPv6 deployment, PINOT converts an encrypted IPv4 packet to an IPv6 packet that embeds the encrypted IPv4 address in the IPv6 source address (similar to NAT46 [9]) to (1) carry encryption-related state and (2) ensure response packets can be forwarded back normally. After encryption, the source IP address in a packet becomes meaningless to an adversary, who only knows which Autonomous System (AS) initiated the packet but cannot pinpoint the specific host that sent the packet, nor associate multiple packets as coming from the same host. Our security analysis demonstrates that PINOT is secure against a realistic adversary under practical constraints.

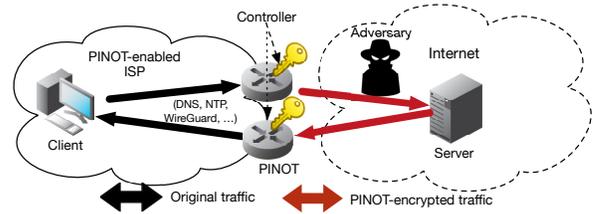

**Figure 1: PINOT setup.**

**Compared to NATs and SPINE.** In contrast to stateful NATs [8], PINOT is stateless for the connectionless protocols that we target, which makes it scalable and also allows it to handle asymmetric routing. With a software controller distributing the per-AS secret keys, an AS can deploy PINOT at multiple border points of the trusted network; outgoing traffic can go through any egress points while return traffic can come into any ingress point. Besides, PINOT offers cryptographic protection for IP addresses and better traffic-analysis resilience than conventional NATs [8]. The support for single-AS deployment and ubiquitous IP obfuscation for all users in the trusted network also ease the deployment of PINOT, compared to other anonymity solutions that require multi-AS cooperation ,e.g., SPINE, or user participation [4–6, 11, 12, 24, 37]. Particularly, SPINE cannot perform encryption at line rate for the connectionless protocols that we focus on (see §6 for a detailed discussion).

We showcase that PINOT can protect users' network identities against malicious public services and eavesdroppers in DNS, NTP, and WireGuard VPN protocols. Because of the design of WireGuard, a WireGuard session, regardless of the protocols underneath, would not be disrupted by PINOT even if every WireGuard packet has a different source IP address [15]. The combination of WireGuard and PINOT can be a useful building block for bootstrapping advanced privacy-enhancing techniques.

We implement a prototype of PINOT on a Tofino switch and deploy it in a campus network to forward DNS, NTP, and WireGuard traffic. We show PINOT is feasible as it can correctly encrypt, decrypt, and forward packets. While previous works have implemented cryptographic algorithms on programming data planes using CPUs, SmartNICs, or NetFPGAs [20, 21, 39], to the best of our knowledge, we are the first to implement a working secure and efficient encryption scheme that runs at line rate within Tofino ASICs.

## 2 PINOT PROBLEM DEFINITION

We study how to design a readily-deployable, in-network privacy-enhancing system with minimal performance overhead. As the first step towards a comprehensive in-network anonymity system, we consider a lightweight notion of anonymity, in which the AS where PINOT is deployed is trusted and can observe both ends of a communication session. We only consider protecting the user's IP address against widely-used public services that are running connectionless protocols, namely, DNS, NTP, and WireGuard. In this section, we elaborate on motivating use cases, design goals, threat model, and hardware resource constraints.

### 2.1 Connectionless use cases

There are numerous privacy threats related to IP address leaks:

**DNS.** DNS recursive resolvers are typically operated by the users' Internet service provider or third-party services (e.g., Google and Cloudflare). Since DNS requests and responses are in cleartext, DNS recursive resolvers or any on-path eavesdroppers can learn the IP address of a user and the domain the user visited. Such information can be further used for inferring users' browsing behaviors, fingerprinting and tracking users, or deanonymizing Tor users [19, 22, 45].

**NTP.** NTP is widely used for synchronizing the clocks of computer systems. A scanner may exploit NTP to discover active IPv6 and IPv4 hosts and conduct unauthorized scans. One such example is shodan.io [40], who contributes its own NTP servers to a public NTP server pool, collects IP addresses of active NTP clients [18, 43], and performs vulnerability scans. The vulnerability information of hosts is publicly-available and can easily be gathered by attackers during reconnaissance, posing security threats to the scanned hosts.

**VPN.** There are privacy concerns with third-party VPN providers. A recent report shows that 25 out of 123 VPN services collect client IP addresses [31]. The IP information could be sold to advertising companies to facilitate delivering of personalized ad content (i.e., IP targeting ads) [17]. Besides, VPN services may unintentionally leak clients' IP information [34]. We focus on WireGuard, a UDP-based, connectionless VPN protocol being added to the Linux kernel [15]. WireGuard leverages a special mechanism to achieve good IP mobility: it assumes a peer's public IP address can change frequently, and maintains a peer address table to record the source IP address in the *latest* packet received from the peer for future commutations. This mechanism enables us to obfuscate peer (user) IP addresses in WireGuard traffic without disrupting connectivity.

### 2.2 Design goals

Motivated by the above use cases, we seek to design a system that achieves the following privacy properties:

**Sender anonymity.** With sender anonymity, an adversary cannot discover the identity (IP address) of the client (sender). We do not try to hide the server's identity or the Autonomous System (AS) of the client.

**Packet unlinkability.** We define packet unlinkability for connectionless protocols as: given a set of packets, the adversary cannot determine whether the packets are associated with the same client based on observed IP addresses. This property helps protect users against traffic-analysis attacks or user tracking. *Associating packets to clients using non-IP information is beyond the scope of this paper.*

Additionally, we want to achieve two operational goals:

**Low deployment barriers.** The solution should be readily deployable without modifications to existing Internet infrastructure and protocols, or running special client-side software (i.e., no involvement of end-users).[1] In addition, the solution should be able to provide ubiquitous privacy protection for a set of users.

**Low performance overhead.** We want our solution to process network traffic at hardware switch rates. Therefore, we need to minimize the overhead introduced by cryptographic operations, and keep as little per-packet/per-flow state as possible (or, better yet, no state at all).

### 2.3 Threat model and assumptions

As shown in Figure 1, we assume an *unmodified* client communicates with a server through a trusted entity (an enterprise network or an ISP). The trusted entity and the server should have both IPv4 and IPv6 connectivity. The goal of an adversary is to recover the actual client IP addresses of network traffic it observes, given the contents of the packets that it can see. We also assume that non-IP fields in packets are appropriately protected.

---
[1]Client-side software could enhance privacy protection further, e.g., Tor browser mitigates many application-layer privacy leaks.

We consider two types of attackers: passive and active. The passive adversary could be the remote server, or any network element between the trusted network and the server. The adversary may be an AS; it could also be any intermediate network point such as an Internet exchange point (IXP) or even simply a link. An active adversary may control a few hosts in the trusted network, and is able to send packets with arbitrary spoofed source IP addresses. We assume, however, that an active adversary cannot observe other users' traffic in the trusted network.

Finally, we do not yet consider implementation-specific attacks, such as bugs in the implementation and bias in the random number generators offered by hardware switches.

## 2.4 Hardware resource constraints

High-speed programmable switches can facilitate achieving the performance goal in §2.2. Programmable switches deployed at the border of the trusted network can process terabytes of traffic per second. Nevertheless, to build a system with the desired privacy properties, we need to work carefully within the hardware resource constraints.

In a programmable switch, the packet-processing pipeline is divided into multiple *stages*. Each stage only allows a limited number of table lookups, and mathematical and logical operations. A program running in the data plane can process traffic at line rate only if it can "fit" into the switch ASIC—that is, if the program only requires the packet to go through the pipeline once. Given the limited stages in commodity programmable switches, fitting standard cryptographic algorithms into the switch, if possible, is extremely challenging (see §6 for more discussions).

## 3 PINOT DESIGN

PINOT consists of a software controller for key management, and a data-plane program that performs IP address encryption/decryption at hardware switch rates using a lightweight yet secure cipher [2]. We use the IPv6 address encoding technique to avoid maintaining any per-flow or per-packet state, similar to NAT46 [9]. With the controller distributing the per-network encryption keys, PINOT can be deployed at multiple border points of the trusted network and naturally handles asymmetric routing. Next, we discuss the encryption scheme and IPv6 encoding.

### 3.1 Efficient encryption in the data plane

Standard encryption algorithms such as AES are too complex to implement in a single pass through the switch ASIC. Performing multiple passes over the packets would cause significant performance degradation, making line-rate encryption infeasible. Therefore, we look for a lightweight and secure cipher that can fit into switch ASICs. After exploring various options, we found that the two-round Even-Mansour (2EM) scheme [2] satisfies our requirements. 2EM can be implemented using table lookups and XORs, avoiding complex cryptographic computations such as hashing in the data plane. With careful code optimization (e.g., rearranging actions using P4 compiler macros to maximize the number of parallel actions in each stage), 2EM can encrypt a packet in a single pass through the packet processing pipeline, yet is secure against a computationally bounded adversary in practice.

Our cipher encrypts a $n$-bit message $M$ by computing:

$$E(M) = P_2(P_1(M \oplus k_0) \oplus k_1) \oplus k_2 \quad (1)$$

where $k_0, k_1$, and $k_2$ are n-bit *independent* encryption keys to thwart attacks that exploit key relation, and $P_1$ and $P_2$ are *independent* permutations over *n*-bit strings, which can be implemented as substitution-permutation networks (SPN) [44]. This construction has been proven to be secure up to $2^{\frac{2n}{3}}$ queries under adaptive chosen-plaintext and ciphertext adversaries [7, 26]. For encrypting 32-bit IPv4 addresses ($n = 32$), 2EM is only secure against 2.6 million queries, i.e., an adversary can recover keys or plaintexts with high probability after knowing 2.6 M plaintext-ciphertext pairs using efficient attacks [7, 30]. To improve the security of PINOT, we adopt two approaches:

• *Random padding to increase message size*: We append a *l*-bit random string to an IP address to extend the length of encryption input. The use of random padding improves the security of the encryption as $n$ becomes larger ($l + 32$). For $l = 32$, our cipher is secure against about 7 trillion queries. Random padding also makes the encryption non-deterministic, i.e., encrypting a given IP addresses multiple times will produce different ciphertexts. This is desirable for achieving packet unlinkability.

• *Key rotation to limit the number of encryptions under given keys*: We update the *key set* (i.e., the three encryption keys) being used for encryption every *t* seconds. Key rotation limits the number of plaintext-ciphertext pairs an adversary can collect for given keys to reduce the attack success probability, and minimizes the damage caused by compromised encryption keys, as keys expire after at most *t* seconds.[2] One potential issue caused by key rotation is inconsistent keys during encryption and decryption, i.e., the key set may be updated when the packets from the server are still in transit. To address this issue, we maintain three versions of key sets, and rotate the key sets using the algorithm proposed in SPINE [11].

A realistic adversary can only perform a reasonable amount of computation (i.e., computationally bounded) under limited memory resources. Given that the adversary might not be able to send more than 1 trillion packets per second (approximately 2.4 Pbps even if assuming the average IPv4 packet size is 300 bytes [3]), the best known practical attacks against 2EM, which are chosen-plaintext attacks, require more than $2^{89}$ bits of memory for $n = 64$ [13]. This is infeasible in practice. In addition, generating a large volume of traffic from a few hosts, if possible, can easily be flagged as DDoS attacks. The adversary may also target a specific IP address, spoofs this IP address, and collects all the possible encrypted IPv6 addresses to build a cipertext table in advance. If it finds a packet's encrypted source IPv6 address in the table, the adversary knows the packet was sent from the target IP address. However, to have a high successful rate, such attacks require the adversary to generate about 10 Tb traffic from a single IP address to build the cipertext table for a given key set, which can also easily be detected. In fact, the adversary cannot perform the best known attacks and the targeted attacks in a majority of networks because of a lack of support for address spoofing (i.e., choosing plaintexts), according to the Spoofer project [27]. Adversaries are also limited in terms of performing chosen-ciphertext attacks because they cannot see the decrypted addresses: PINOT may forward a tampered packet based on the decrypted address, or may drop the packet because it does not recognize the decrypted address. See Appendix §A.1 for a detailed security analysis.

---
[2]We can also rotate the permutations.

## 3.2 Translation from IPv4 to IPv6

Encrypting a 32-bit IPv4 address will produce a ciphertext of $32 + l$ bits ($l$ is the length of the random padding), which cannot be used as a valid IPv4 address for routing. The whole ciphertext also needs to be stored somewhere for decryption. To ensure return traffic can be correctly routed back based on the encrypted IP addresses and to avoid maintaining any state, we transform an IPv4 packet into an IPv6 packet when the packet leaves the trusted network. The IPv6 packet encodes the encrypted IPv4 address in its IPv6 source address field, as shown in Figure 2. The highest $d$ bits of a transformed IPv6 source address are the IPv6 network prefix reserved by the trusted network, the lowest $32 + l$ bits contain the encrypted IPv4 source address, and the remaining bits are used for storing encryption metadata such as the key set version number. The values in the IPv4 header fields that have corresponding IPv6 header fields are preserved.

The IPv4 destination addresses are replaced with their IPv6 counterparts. Recall that we only consider services with both IPv4 and IPv6 addresses. Therefore, we can perform DNS lookups in advance to get the IPv4 and IPv6 addresses of public servers of interest from their DNS A and AAAA records, respectively, and store the IPv4/IPv6 address pairs in a lookup table for later use.

We use the key version number in the IPv6 destination address of a return (IPv6) packet to locate the key set for decryption. The return packet is converted back to an IPv4 packet and forwarded based on the decrypted address.

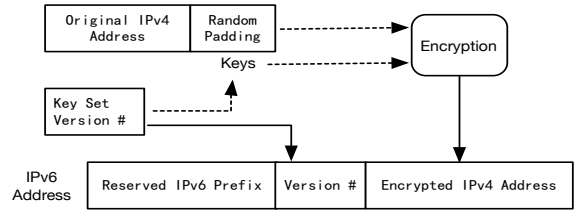

Figure 2: PINOT IPv6 source address encoding.

## 3.3 PINOT in the wild

PINOT can improve user privacy for DNS, NTP, and WireGuard VPN services. PINOT moves trust from third-party servers to the trusted network so that no party outside the trusted network can see the real originating IP address of a packet. Unlike NAT that assigns a user the same IP for a certain time period, PINOT assigns a random IP to every packet, which is more effective at defeating user tracking based on traffic patterns [22]. In addition, PINOT makes unsolicited scanning harder, as the mapping between the IP address and end-user device changes frequently. Finally, PINOT can perform IP address obfuscation at line rate for all users in the trusted network without cooperation from users, the destinations or any other AS (other than the trusted AS), making it less prone to human errors and issues caused by extra latency (e.g., inaccurate time synchronization in NTP).

## 4 IMPLEMENTATION

PINOT consists of a software controller for key distribution and rotation, and a P4 data-plane program for encryption. The controller, which can run on a dedicated host or in the control plane of a programmable switch, generates three 64-bit encryption keys using the Python urandom function. It uses grpc to communicate with the data plane to update the keys and the key version number. In our evaluation, the keys are updated every five seconds.

The two-bit version number is stored along with the port forwarding (i.e., switch ingress port to egress port) information in a forwarding table. For IPv6/IPv4 packet transformation, PINOT also maintains two address mapping tables (*IP4to6* and *IP6to4*) that store the corresponding IPv6 address of an IPv4 address, and vice versa, for the servers.

PINOT generates random paddings via the Random external function in the Tofino switch, and uses substitution-permutation networks for permutation [44]. To permute a 64-bit input, PINOT first performs substitution using 8-bit substitution boxes (S-boxes) to substitute every byte of the input with another byte, and then applies a 64-bit straight permutation box (P-box) to shuffle the bits of the S-box output. We currently use the static S-box in the AES standard and randomly shuffle the bits in the P-box; we plan to generate S-boxes and P-boxes dynamically and update them periodically in the future.

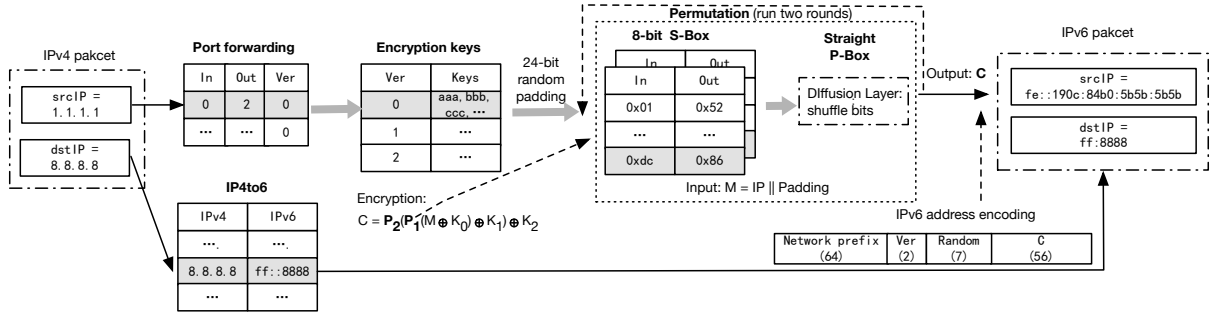

Figure 3: IPv4 source address encryption in 56-bit PINOT.

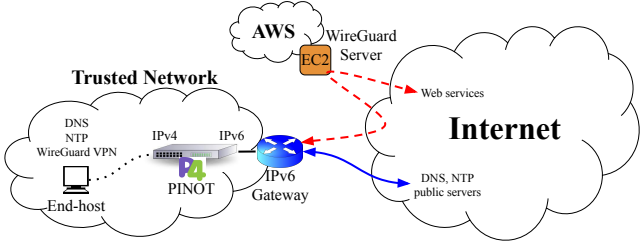

Figure 4: PINOT deployment setup.

We implement two different PINOT prototypes: a 56-bit version and a 64-bit version that use 24-bit and 32-bit random paddings (i.e., $l = 24$ and $l = 32$), respectively. We use the 56-bit version for the real-world deployment because we have a /64 IPv6 network allocated; with the 64-bit version, there is no space left in an IPv6 address for the two-bit version number. Figure 3 shows an example of IPv4 source address encryption in the 56-bit PINOT. Even with the 64-bit PINOT, the S-box or inverse S-box tables only take up about 16 KB of memory, and the extra memory used for storing encryption keys and P-boxes are negligible. Such low overhead allows PINOT to store additional encryption keys to encrypt more fields in the packet header. See Appendix §A.2.

## 5 DEPLOYMENT AND EVALUATION

**Wide-area tesbed:** Figure 4 shows the PINOT deployment in our network that connects to the wider Internet. The end-host IPv4 client device is a Linux server with two Intel Xeon E5530 2.4GHz CPUs and 16GB of memory. The PINOT switch sits between our end-host device and the trusted network's border gateway. The switch is a Wedge 100BF-32X switch with a Tofino programmable chip [29], and loads the PINOT P4_16 program. The switch acts as the IPv4 gateway for the client. On the IPv6 side, our network's border gateway allocates a /64 IPv6 subnet to the PINOT switch; PINOT selects IPv6 addresses in this subnet that are used by the IPv4 client when communicating with the Internet.

The client's DNS, NTP, and WireGuard traffic traverse the PINOT switch and the IPv6 border gateway to reach public servers on the Internet. The PINOT switch automatically translates a target server address to an IPv6 equivalent and vice versa for the response using pre-installed rules. For the WireGuard VPN experiment, we set up a WireGuard forwarding server on an AWS EC2 t2.micro instance. Our client's WireGuard traffic traverses this server to reach web services on the Internet.

We test PINOT on DNS, NTP, and WireGuard traffic in our evaluation. From the lists of 11,884 public DNS resolvers [35] and 223 NTP servers [28], we found 374 DNS resolvers and 145 NTP servers that have both IPv4 and IPV6 addresses thus can answer IPv4 or IPv6 queries correctly. The IPv4/v6 address pairs of these servers are stored in the PINOT switch's IP4to6 and IP6to4 tables. We omit the evaluation results for NTP due to space considerations; note that the case is similar to DNS as both are single-packet protocols.

**Local testbed:** For the performance evaluation (Section 5.2), we set up a WireGuard tunnel in a local network. This is to minimize the effect of network conditions on throughput. The two machines (Intel Xeon 2.2GHz CPU, 96 GB Memory) running WireGuard are connected to a Tofino Edge-Core Wedge 100BF-32X switch using 40GbE links. One machine is used as the client and the other one serves as the server.

### 5.1 Feasibility

We evaluate if PINOT correctly encrypts, decrypts, and forwards DNS and WireGuard traffic.

**DNS.** We use dig to send DNS queries for A records of ten unique domains randomly selected from the top 1 million domains to each resolver (totalling 3,740 queries), using IPv4 and IPv6 networks. There are 3,387 *consistent* queries, i.e., returning the exact same A records in both settings; the inconsistent responses are caused by either DNS load balancing or resolver-side errors (misconfiguration, etc.). We replay the consistent queries with PINOT running and find all returned A records are consistent with the results that use IPv4 or IPv6 networks.

**WireGuard.** We download 100 randomly selected files with varying sizes (1 KB to 10 GB) from two websites [32, 41]. All files download successfully through WireGuard and PINOT, and the SHA1 hash of every file matches that of equivalent downloads directly using the IPv4 network.

Overall, we conclude that per-packet encryption does not affect the normal use of DNS and WireGuard.

## 5.2 Performance

Though encryption and decryption can be performed at switch hardware rates (e.g., 3.2 Tb/s on our switch), there are two potential sources of overhead:

**Latency introduced by IPv6 routing.** The routing paths taken by IPv6 packets and IPv4 packets could be different, affecting latency. We test 1,000 DNS queries and examine the query time. Though using IPv6 may add up to 30 ms of delay in a query, PINOT does not introduce additional latency in 97% of the cases.

**WireGuard throughput degradation caused by address table update.** WireGuard may update its peer address table more frequently to keep track of the ever-changing peer IP addresses, which could affect its throughput. To minimize the effect of other network conditions, we use the local testbed described earlier in this section. We conduct throughput tests with the switch running PINOT that performs per-packet encryption (refer to this setting as PINOT-pp), and compare the results to two baseline settings: (1) NAT: The switch acts as an NAT46 and simply converts IPv4 addresses to fixed IPv6 addresses. (2) PINOT-$t$: The switch runs a modified version of PINOT that uses the same padding for $t$ seconds. All the packets sent from the same client during that $t$ seconds will have the same encrypted source address.

We run iperf TCP throughput tests (without optimization) for 300 seconds and collect the throughput reported by iperf every second. As shown in Figure 5, the throughput decreases as the address table update frequency increases. The average throughput drops from 8.3 Gbps (without address table update) to 6.7 Gbps (with per-packet address table update). The performance bottleneck is the server's CPU rather than PINOT.[3] If sender anonymity is the priority concern, one may lessen this degradation by using fixed random padding for a batch of packets, as in PINOT-5$s$. In fact, WireGuard includes public peer identity information that can be used to associate packets to WireGuard sessions in packet payloads. Future work on improving WireGuard against traffic analysis attacks may consider mitigating application layer leaks with QUIC's payload encryption mechanism.

---

[3]iperf CPU utilization is near 100% in all the settings.

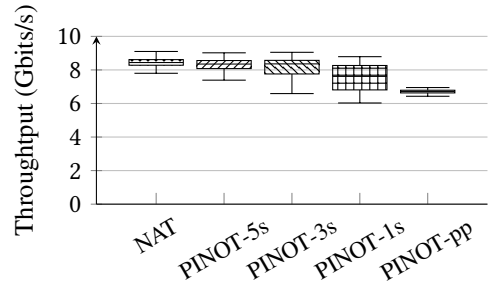

Figure 5: WireGuard throughput tests.

## 6 RELATED WORK

**Cryptographic algorithms in programmable data planes.** The ability to support cryptographic algorithms in programmable hardware is important for offloading security and privacy applications to data planes. Previous works mostly focus on using switch CPUs, SmartNICs, or NetFPGAs to implement cryptographic algorithms, but these approaches may have performance, scalability, or compatibility issues [20, 21, 39]. Very few studies have examined using switch ASICs for cryptographic operations. SPINE implements a prototype of SipHash for the BMv2 software model [11]. However, unlike in software, SPINE likely needs at least three passes of the packet on a hardware switch due to resource constraints, degrading the throughput by a factor of three. Similarly, P4-AES [1] requires at least two more passes of a packet on hardware switches. In contrast to SPINE and P4-AES, PINOT is able to fit into switch ASICs and encrypt source IP addresses using a single pass of the packet processing pipeline on commercial off-the-shelf programmable switches.

**Hiding user IP addresses.** Network-layer anonymity systems, such as LAP [24], Dovetail [37], HORNET [4], PHI [6], and TARANET [5], typically require multiple ASes along an end-to-end path to cooperate in the protocol and end-users to run specialized software. The involvement of end-users not only further raises deployment barriers, but also introduces human errors [33] that cause privacy failures. None of the systems were implemented on programmable hardware.

Address Hiding Protocol (AHP) [36] and SPINE [11] can conceal users' IP addresses without user participation. In AHP, a trusted network assigns a random IP address to a user from its own IPv4 address space, which poses security issues for small networks. Besides, AHP does not provide packet unlinkability. SPINE encrypts the IP address in every packet using a programmable switch, but requires an additional trusted network to decrypt every packet.

For DNS, DNS-over-HTTPS (DoH), DNS-over-TLS (DoT), DNSCrypt [14, 23, 25], or VPNs can protect users' IP addresses. However, they put trust in third-party servers, which could become a single point of privacy failure. Oblivious

DNS [38] hides IP addresses from third-party resolvers, but requires modifications to clients and infrastructure. In contrast to known anonymity solutions, PINOT does not require cooperation from end-users nor third-party services.

## 7 CONCLUSION

PINOT is a lightweight in-network anonymity solution that hides users' IP addresses from downstream ASes and destination servers. Utilizing an efficient and secure encryption scheme, PINOT can encrypt IP addresses at hardware switch rates. In contrast to known anonymity solutions, PINOT has a low barrier to deployment, because it requires no cooperation from end-users or any ASes other than the trusted network where it is deployed. We implemented and deployed a prototype of PINOT, and demonstrated PINOT is feasible for improving user privacy in DNS, NTP, and WireGuard VPN protocols.

## A APPENDIX
## A.1 Security analysis

The security of 2EM depends on the plaintext size $n$ ($32 + l$). With a larger IPv6 network (i.e., a short network prefix $d$), PINOT can append more random bits to the original IPv4 source address to produce a longer input. To facilitate discussion, we fix $n = 64$ (i.e., $l = 32$). Note that $2^{\frac{2n}{3}}$ is the lower bound obtained in the information-theoretic model [7, 26]. There is a significant gap between this lower bound and the complexity of the best-known attacks in the computational model. We only consider the best-known attacks for certain types of (computationally bounded) adversaries in this section.

**Passive adversary.** A passive adversary can only perform the trivial attacks, i.e., an exhaustive key search, which requires brute forcing in the space of $2^{3n}$ or $2^{192}$ keys. Such brute-force attacks are clearly infeasible for computationally bounded adversaries.

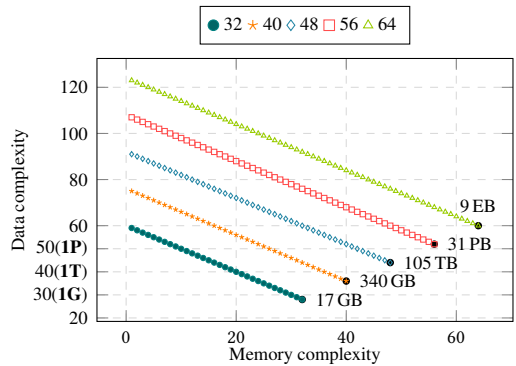

Figure 6: Data complexity vs memory complexity of the key recovery attacks under varied $n$ (32, 40, 48, 56 and 64 bits). The memory complexity increases from $2^1$ to $2^n$. X-axis and Y-axis are in log2 scale. Data complexity indicates the number of plaintext -ciphertext pairs (i.e., encrypted packets) an adversary needs to collect.

**Known/chosen-plaintext adversary.** Non-trivial attacks usually need to consider three important factors: data complexity, memory complexity, and time complexity, where data complexity is the number of ciphertext-plaintext pairs they need to collect, and memory complexity is the number of memory units ($n$-bit blocks) required during attacks. A known-plaintext adversary may trade memory complexity for data complexity using the man-in-the-middle (MITM) attacks proposed by Andrey *et al.* [2]. A MITM attack only requires knowing a small number of ciphertext-plaintext pairs (e.g., 2); however, its memory complexity is $2^n$, i.e., more than 147 exabytes of memory is required for storing a precomputed table.

The best *known* attacks against 2EM with independent keys and permutations are the key recovery attacks proposed by Dinur *et al.* [13], aiming to lower time complexity. The key recovery attacks also require precomputing a big table, and the memory complexity $M$ and data complexity $D$ approximately satisfy $D = 2^{2(n-1)}/(4M)$. The memory-data trade-offs are shown in Figure 6. Indeed, a powerful adversary may be able and be willing to prepare a large amount of memory. However, *practical attacks are hard.*, as the adversary must obey rate limits. In practice, even a large ISP may not see more than 1 T packets per second (approximately 2.4 Pbps assuming the average IPv4 packet size is 300 bytes). Under this constraint, the adversary still needs to prepare $2^{89}$ bits of memory for $n = 64$, and more memory if targeting a lower packet rate. The use of random padding also makes attacks harder because the adversary can only choose or know partial plaintexts.

Besides, the adversary cannot perform chosen-plaintext attacks in a majority of networks or ASes because of a lack of capability on source address spoofing (i.e., choosing plaintexts). The Spoofer project examined more than 7 K /24 networks and found about 85% of them implement certain mechanisms (e.g. Source Address Validation) to filter outbound spoofed-source packets [27]. For an active adversary, generating a large volume of traffic from a few hosts, if possible, can easily be flagged as DDoS attacks.

**Chosen-ciphertext adversary.** Our encryption scheme does not provide malleability so the adversary might be able to manipulate the destination addresses in the return traffic. PINOT may forward a tampered packet based on the decrypted address, or may drop the packet because it does not recognize the decrypted address. In either case the adversary cannot see the decrypted address, and therefore is limited in terms of performing chosen-ciphertext attacks.

**2EM alternatives.** It is possible to fit the standard ciphers with reduced rounds into the data plane, e.g., 2-round AES and 2-round DES. However, the adversary may break these schemes with low data complexity attacks by exploiting the relation between round keys and algebraic properties of the algorithms [10, 16].

In our setting, a cipher is considered as secure if breaking it requires attacks with high data complexity. Besides, the encrypting and decrypting parties are the same in PINOT, so we do not need to consider key distribution and can store key materials of large sizes. A good alternative to 2EM should use independent round keys to prevent the adversary from exploiting its key schedule. We believe there are other ciphers can be used in lieu of 2EM, and leave exploring 2EM alternatives as future work.

## A.2 Obfuscating other IP header fields

We have extended our prototypes to support port encryption by adding a *one-time pad* table that stores random 16-bit one-time pads. PINOT uses the first 16 bit of the generated random padding as a key to fetch the corresponding one-time pad in the one-time pad table, and XOR the one-time pad with the source port in an IPv4 packet. Fetching one-time pads can be done in parallel with IP address encryption, requiring no additional stages. In fact, we can get multiple one-time pads at the same time in one table lookup, and XOR them with different header fields.

## A.3 PINOT for IPv6 networks

We can also use PINOT to protect certain IPv6 networks. For instance, the trusted entity has a /64 IPv6 network, and reserves a /96 network for PINOT. We call the 64 to 96 bits in an IPv6 source address *subnet ID*, which is static in this case. We can use PINOT to encrypt the lowest 32 bits of an IPv6 source address with at most 30 bits of random padding, and use subnet ID to carry the encryption meta. During decryption, PINOT replaces the subnet ID part of an IPv6 address with the original, static subnet ID.